\documentclass[12pt]{iopart}
\usepackage{graphicx}

\usepackage{amssymb}
\begin{document}

\title[A Monte Carlo Calculation of the Pionium Break-up Probability]%
{A Monte Carlo Calculation of the Pionium Break-up Probability with
  Different Sets of Pionium Target Cross Sections}

\author{C Santamarina$^\dagger$, M Schumann$^\ddagger$, L G
  Afanasyev$^\S$ and T Heim$^\ddagger$}
\address{$\dagger$ Institut f\"ur Physik, Universit\"at Basel,
  4056 Basel, Switzerland}
\address{$\ddagger$ Institut f\"ur Theoretische Physik, Universit\"at Basel,
  4056 Basel, Switzerland}
\address{$\S$ Joint Institute for Nuclear Research, 141980 Dubna,
Moscow Region, Russia}

\begin{abstract}
Chiral Perturbation Theory predicts the lifetime of pionium, a
hydrogen-like $\pi^+ \pi^-$ atom, to better than 3\% precision. The goal of the
DIRAC experiment at CERN is to obtain and check this value
experimentally by measuring the break-up probability of pionium in a
target. In order to accurately measure the lifetime one needs to know
the relationship between the break-up probability and lifetime to a
$1\%$ accuracy. We have obtained this dependence by modeling the
evolution of pionic atoms in the target using Monte Carlo methods. The
model relies on the computation of the pionium--target atom interaction
cross sections. Three different sets of pionium--target cross sections
with varying degrees of complexity were used: from the simplest first order
Born approximation involving only the electrostatic interaction to a
more advanced approach taking into account multi-photon exchanges
and relativistic effects.  We conclude that in order to
obtain the pionium lifetime to 1\% accuracy from 
the break-up probability, the pionium--target cross sections must be known
with the same accuracy for the low excited bound states of the pionic atom.
This result has been achieved, for low $Z$ targets,
with the two most precise cross section sets. For large $Z$ targets
only the set accounting for multiphoton exchange satisfies the condition.

\end{abstract}


\submitto{\JPB}

\maketitle

\section{Introduction}

Pionium is the hydrogen-like electromagnetic bound state of a
$\pi^+ \pi^-$ pair. Its lifetime is determined by the strong
$\pi^+ \pi^-
\rightarrow \pi^0 \pi^0$ annihilation in the low relative momentum regime
where Chiral Perturbation Theory applies. The theory predicts the
pionium lifetime value of $(2.9 \pm 0.1)\times 10^{-15}\mbox{ s}$~\cite{cola}.
Experimentally, this value is being measured in
a model-independent way by the DIRAC experiment at CERN~\cite{prop}.

Under the experimental conditions of DIRAC the pionic atoms are created
in the inelastic scattering between $24\;\textrm{GeV}/c$ protons, supplied by
the PS at CERN, and the nuclei of the target~\cite{neme,gorc}, a chemically pure
material of well-determined thickness.

A pionic atom propagating at relativistic speed inside the target 
interacts mainly electromagnetically with the target
atoms. The electromagnetic cross section is on the order of a megabarn.
These interactions can lead either to a transition between two
bound states of pionium or to a dissociation (break-up). The scattering
with the target atoms competes with the process of annihilation since
a shorter pionium lifetime leads to a smaller break-up
probability.

The main thrust of DIRAC's experimental technique is the detection
of $\pi^+ \pi^-$ pairs with very low relative momentum. From the
low-momentum part of the spectrum, the break-up probability of pionium
is then determined.

The goal of this study is to establish the theoretical dependence of
the break-up probability on the lifetime with a very high accuracy.
We modeled the dynamics of pionium in different targets
taking into account a large number of atomic shells that become
populated as the pionic atom evolves in the target.

We have performed the calculations by using three sets of pionium
target interaction cross sections. The original studies by Afanasyev
and Tarasov~\cite{afan} made use of the Born approximation and pure
electrostatic interaction. In our work we have applied new corrected
cross sections that
take into account relativistic effects~\cite{heim} and
the multiphoton exchanges~\cite{schu}. In addition, these cross
section sets  include a more accurate description of the target
atomic form factor. As an error in the break-up probability translates
directly to an error in lifetime, we have checked whether the results
obtained in the new
calculations significantly deviate from the previous ones. We
conclude that the magnetic and relativistic corrections together with
the target form factor choice are non-negligible mainly for the small
$Z$ targets, whereas the multiphoton exchange should be considered
for the large $Z$ ones.

\section{Monte Carlo Simulation of Pionium in the Target}

As we have noted in the introduction,
to determine the pionium lifetime, the  experiment DIRAC needs a
precise theoretical calculation of the break-up probability
$P_{\mathrm{br}}$ of pionium due to the electromagnetic interaction with target atoms.
This calculation can be done by means of a Monte Carlo transport code
that simulates the evolution of pionium from its creation to either its
annihilation or its break-up under the given experimental conditions.

\subsection{Pionium Production}
\label{creation}

The pionic atoms are formed as a consequence of the Coulomb final
state interaction of two oppositely charged pions.
These pions are created in an inelastic collision of a
$24\;\textrm{GeV}/c$ proton
and one of the target nuclei.
The pionic atom is described by six quantities,
accounting for the six degrees of freedom of a two body system. A
particularly convenient choice of these quantities is given by the
laboratory momentum of the center of mass of the atom $\vec{P}$ and
the quantum numbers of the created bound state $n$, $l$, and $m$
in the spherical coordinate
representation~\footnote{Also parabolic quantum numbers have been used
  elsewhere~\cite{afan2}.}.

The probability of pionium being created is given by~\cite{neme}
\begin{equation}
\frac{\rmd\sigma^A_{nlm}}{\rmd\vec{P}} = 
(2\pi)^3 \left|\psi_{nlm}(0)
\right|^2\vphantom{\Bigg|_{\vec p}}
\frac{E}{M}
\frac{\rmd\sigma^0_{s}}{\rmd\vec{p}\,\rmd\vec{q}}
\Bigg|_{\vec{p}=\vec{q}=\vec{P}/2}
.
\end{equation}
The two terms on the right-hand side of the equation illustrate the
final state interaction mechanism. The rightmost factor is the 
doubly inclusive cross section of
$\pi^+$ and $\pi^-$ pairs at equal momenta ($\vec{p}=\vec{q}$) without considering
the final state interaction, as indicated by the superscript $0$.
The
subscript $s$ means that only pions
created from direct hadronic processes and decays of resonances with a very short lifetime are
considered, because the Coulomb interaction of pions from
long-lived sources (e.g.\ $\eta$, $K^0_S$ and $\Lambda$) is
negligible and hence they do not contribute to pionic atoms production. 
The effect of the final state Coulomb interaction is to create
a bound state with quantum numbers $n$, $l$, and $m$; it 
is given by the squared wave function at
the origin. 

The doubly inclusive cross section can be obtained from the direct
measurements of time correlated $\pi^+ \pi^-$ pairs in DIRAC, according
to the following reasoning:
\begin{itemize}
\item The final state Coulomb interaction for short-lived sources
  is given, as in the case of the creation of a bound state, by a
  multiplicative factor depending only on $Q$, the magnitude of
  the relative momentum between the two pions. This is the so-called
  Coulomb or Gamow factor~\cite{lan} 
  \begin{equation}
    \label{ccfac}
    \frac{\rmd\sigma_s}{\rmd\vec{p}\,\rmd\vec{q}} =
    A_C(Q) \frac{\rmd\sigma^0_s}{\rmd\vec{p}\,\rmd\vec{q}}\; ; \qquad 
    A_C(Q) = \frac{2\pi M_{\pi} \alpha /Q}{1-\rme^{-2 \pi 
    M_{\pi} \alpha /Q}}\; ,
  \end{equation}
  where $\alpha$ is the fine structure constant.
\item The contribution to the doubly inclusive cross section of pairs
   containing at least one pion from a long-lived source,
  $\omega_l (\vec{P})$, can be calculated with a hadron physics Monte
  Carlo simulation. In our case we have used
  FRITIOF6~\cite{friti}. This function has been shown to depend only
  on $P$~\cite{afan0}, the magnitude of the total momentum of the pion pair. 
  Taking this into account together
  with~\eref{ccfac} we find
  \begin{equation}
    \frac{\rmd\sigma}{\rmd\vec{p}\,\rmd\vec{q}} =
    \frac{\rmd\sigma_s}{\rmd\vec{p}\,\rmd\vec{q}} + 
    \frac{\rmd\sigma_l}{\rmd\vec{p}\,\rmd\vec{q}}
    = A_C (Q) \frac{\rmd\sigma_s^0}{\rmd\vec{p}\,\rmd\vec{q}}
    + \omega_l (P) \frac{\rmd\sigma}{\rmd\vec{p}\,\rmd\vec{q}}\; ,
  \end{equation}
  thus relating $\sigma$ and $\sigma_s^0$.
\item Finally, we have found that the $\vec{P}$-dependence of the
  doubly inclusive cross section is not correlated to $\vec{Q}$, given that
  $\vec{Q} \ll 30\; \textrm{MeV}/c$.
\end{itemize}

\begin{figure}
\centerline{\includegraphics[width=7cm]{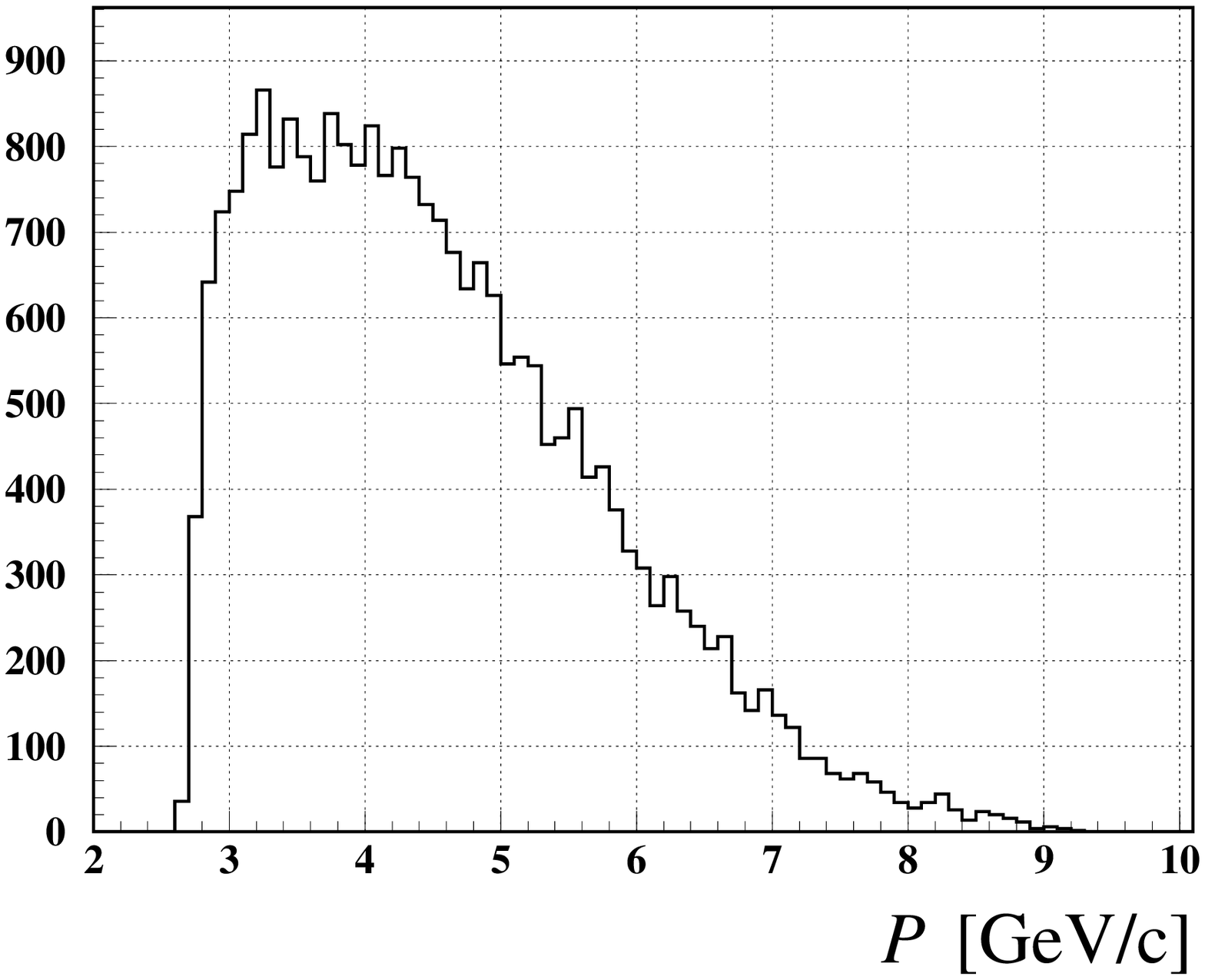}
\includegraphics[width=7cm]{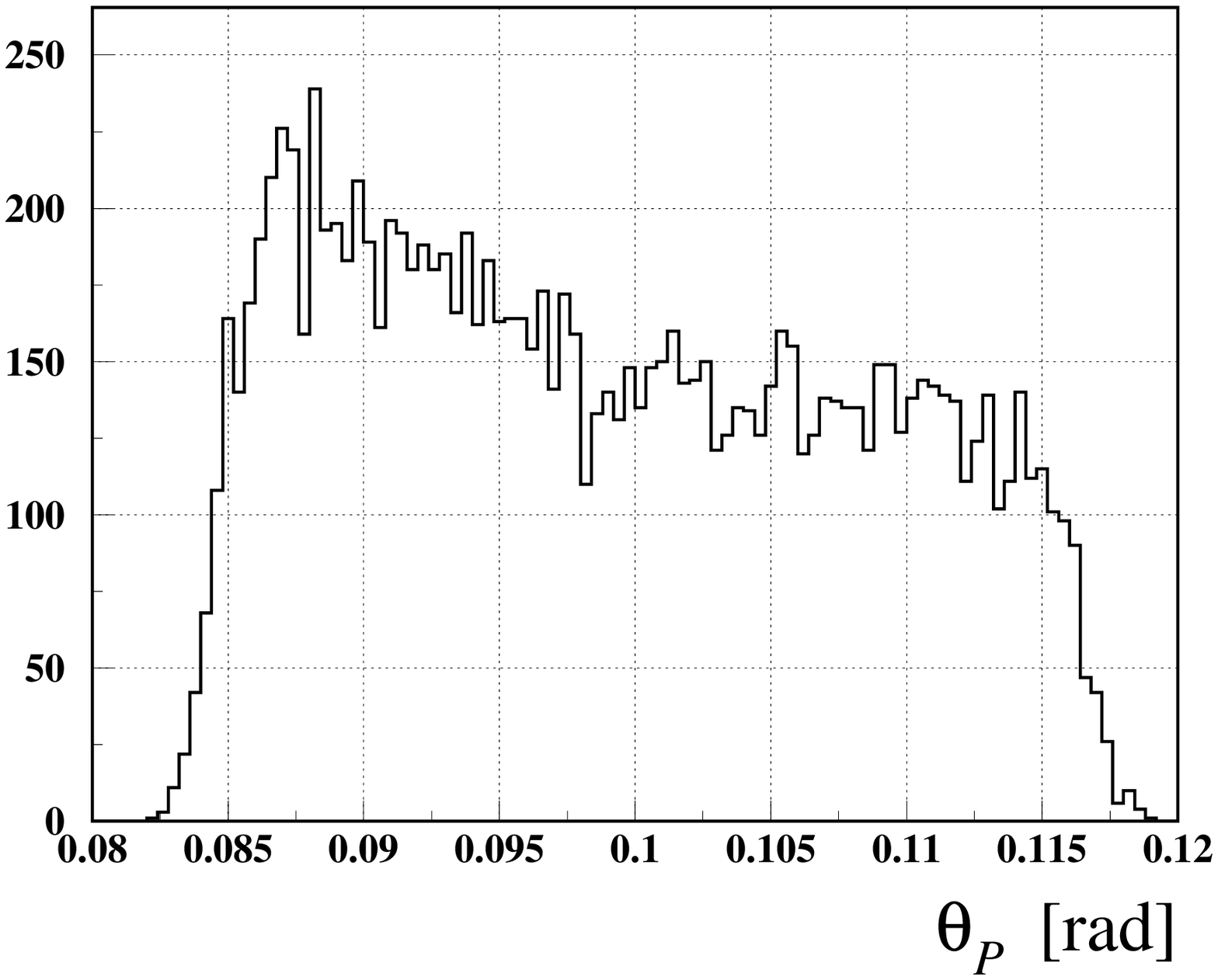}}
\caption{\label{labmom}The left graph shows the laboratory momentum magnitude
distribution and the right graph the angular distribution with respect
to the proton beam axis for low relative momentum $\pi^+\pi^-$ pairs.}
\end{figure}

These findings allow us to relate the $\vec{P}$-dependence of
$\sigma$ and $\sigma_s^0$ by
\begin{equation}
\left. \frac{\rmd\sigma^0_{s}}{\rmd\vec{p}\,\rmd\vec{q}}
\right|_{\vec{p} = \vec{q}=\vec{P}/2} \propto \int_0^{Q \sim 2\; 
\mathrm{MeV}/c}
(1-\omega_l(P))\frac{\rmd\sigma}{\rmd\vec{p}\,\rmd\vec{q}}\, \rmd\vec{Q}\; ,
\end{equation}
where the $\vec{P}$ distribution is obtained from the
direct measurement of the laboratory momentum of low relative momentum
$\pi^+\pi^-$ pairs in DIRAC. In~\fref{labmom} we show the distribution
of the magnitude of the momentum $P$ and the angular distribution
relative to the proton beam axis for low relative momentum $\pi^+\pi^-$ pairs.

The initial quantum number distribution depends on the value of the wave
function at the origin. It has been shown~\cite{kura} that the effect of the strong
interaction between the two pions of the atom significantly modifies
$|\psi_{nlm}(0)|$ in comparison to the pure Coulomb wave function.
However, the ratio between the production rate in different states
has been demonstrated to be kept as for the Coulomb wave functions~\cite{amir}.
Then, considering that the Coulomb functions obey
\begin{equation}
\label{wfat0}
\left|\psi^{(\mathrm{C})}_{nlm}(0)\right|^2 = \left\{
\begin{array}{cl}
0 & \textrm{if $l \neq 0$,} \\
\displaystyle\frac{(\alpha M_{\pi}/2)^3}{\pi n^3} & \textrm{if $l=0$,}
\end{array}
\right.
\end{equation}
we note that only $S$ states are created, according to a $1/n^3$
distribution.

Another quantity to be specified is the position where the
proton--target interaction took place ($\vec{R}$). This is also the
position where the atom was created. Since the target thickness is
chosen much smaller than the nuclear interaction length of the target
material, the $\pi^+\pi^-$ atoms are supposed to be uniformly generated
all across the target thickness. The position in the transverse
coordinates is unimportant, but it can be generated according to the
beam profile characteristics, too.

We have now gathered all the information needed to simulate the creation of
pionium atoms with a center of mass momentum $\vec P$ according to the
experimentally measured momentum and angular distributions, at a position
$\vec R$ uniformly distributed through the target, and in an initial
$S$-wave state with the principal quantum number $n$ distributed according
to a $1/n^3$ distribution.

\subsection{Pionium Annihilation}

Once an atom has been created in its initial state, specified by $\vec{P}$,
$n$, $l$, and $m$, its dynamics are those of a free system that can
either be annihilated, mainly via the $\pi^+ \pi^-\rightarrow \pi^0
\pi^0$ channel, or be electromagnetically scattered by one of the
target atoms.

The strong interaction decay to two neutral
pions determines the lifetime and is related to the $a_0^0-a_0^2$
scattering lengths difference and to the wave function at the origin
by~\cite{gasse}
\begin{equation}
\label{tau}
\frac{1}{\tau_{nlm}} = \frac{16\pi}{9}\frac{\sqrt{M_{\pi}^2-M_{\pi^0}^2
-\frac{1}{4} M_{\pi}^2\,\alpha^2}}{M_{\pi}} (a_0^0 -a_0^2)^2
(1+\delta_{\Gamma}) \left|\psi^{(\mathrm{C})}_{nlm}(0)\right|^2 \; ,
\end{equation}
where $M_{\pi}$ and $M_{\pi^0}$ are the masses of the charged and the
neutral pion, respectively, 
and $\delta_{\Gamma}$ is the correction to next-to-leading order 
($\delta_{\Gamma} = 0.058$) that includes the effect of the strong
interaction between the two pions. 
Using Chiral Perturbation Theory, Colangelo \etal~\cite{cola} have
been able to calculate the most precise value of the scattering
lengths difference to date
($a_0^0 -a_0^2=0.265 \pm 0.004$).
Employing this value in \eref{tau} yields
\begin{equation}
\tau_{100} = (2.9 \pm 0.1)\cdot 10^{-15}\; \textrm{s}.
\end{equation}
Note, however, that due to \eref{wfat0} 
pionium may only decay from $S$ states.
Moreover, the lifetime of any $S$ state is 
related to the lifetime of the ground state, by
\begin{equation}
\tau_{n00} = n^3 \tau.
\end{equation}
For the purpose of simulating pionium in the target, we shall from now on 
refer to $\tau$ as the \emph{pionium lifetime}.

Hence, the probability for a $\pi^+\pi^-$ atom to annihilate per
unit length, after the Lorentz boost transformation to the
laboratory system, is given by 
\begin{equation}
\label{pannih}
p_{nlm}^{\mathrm{anh}} = \frac{1}{\lambda_{nlm}^{\mathrm{anh}}} =
\left\{ \begin{array}{c@{\hspace{1cm}}l}
\displaystyle\frac{2M_{\pi}}{P n^3 \tau} & \textrm{if $l=0$,} \\
0 & \textrm{other cases,}
\end{array}
\right.
\end{equation}
where $\lambda_{nlm}^{\mathrm{anh}}$ is the annihilation mean free path.

\subsection{Electromagnetic Interaction of Pionium with the Target}

The electromagnetic pionium--target interaction of a pionic atom in an
initial $n l m$ bound state can induce a transition to another
$n^{\prime} l^{\prime} m^{\prime}$ bound state. The probability of
such an interaction per unit length is given by
\begin{equation}
\label{pdisc}
p_{nlm}^{n^{\prime} l^{\prime} m^{\prime}} = 
\frac{\rho N_0}{A}\sigma_{nlm}^{n^{\prime} l^{\prime} m^{\prime}}
\end{equation}
where $\rho$ is the target density, $A$ its atomic weight, $N_0$ is
the Avogadro number, and 
$\sigma_{nlm}^{n^{\prime} l^{\prime}  m^{\prime}}$ are the discrete
(bound--bound) transition cross sections.

The break-up mechanism is analogous to the discrete one; the break-up
probability per unit length of an atomic bound state $nlm$ is given by
\begin{equation}
\label{pbreak}
p_{nlm}^{\mathrm{br}} = \frac{1}{\lambda_{nlm}^{\mathrm{br}}}=
\frac{\rho N_0}{A}\sigma_{nlm}^{\mathrm{br}}
\end{equation}
where $\sigma_{nlm}^{\mathrm{br}}$ is the break-up (ionization) cross
section.

Finally, the total cross section gives the probability of an atom to
undergo an electromagnetic interaction and of course 
fulfills
\begin{equation}
\label{sigtot}
\sigma_{nlm}^{\mathrm{em}}=
\sum_{n^{\prime} l^{\prime} m^{\prime}} \sigma_{nlm}^{n^{\prime} l^{\prime}
m^{\prime}} + \sigma_{nlm}^{\mathrm{br}}\; .
\end{equation}
The total probability for a pionic atom to suffer an electromagnetic
collision per unit length is then given by
\begin{equation}
\label{pem}
  p_{nlm}^{\mathrm{em}} = \frac{1}{\lambda_{nlm}^{\mathrm{em}}} = 
  \frac{\rho N_0}{A}\sigma_{nlm}^{\mathrm{em}}\;,
\end{equation}
where $\lambda_{nlm}^{\mathrm{em}}$ is the mean free path before an
electromagnetic interaction takes place. 
Exploiting the completeness of the eigenstates of the Coulomb Hamiltonian the
total electromagnetic cross sections can be  calculated directly~\cite{mrow,afan3} and
not just via~\eref{sigtot}~\footnote{Note, however, that this is strictly
true only within the framework of the sudden
approximation~\cite{heim}.}.

The electromagnetic cross sections have been obtained with different
approaches in~\cite{afan,heim,schu}. We will devote section~\ref{results}
to discussing the different break-up probabilities they lead to.

\begin{figure}
\centerline{\includegraphics[width=7.5cm]{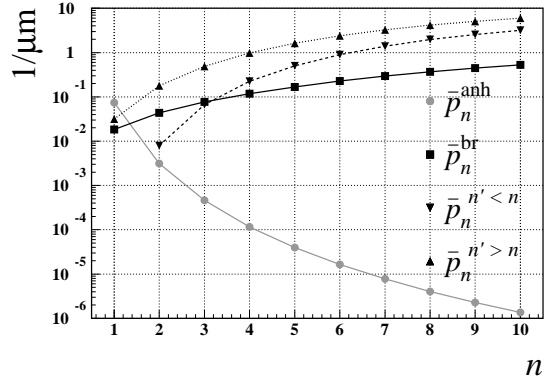}}
\caption{\label{dip} Annihilation, ionization, excitation and de-excitation
  probabilities per unit length, averaged over $l$ and $m$ quantum numbers
  according to equations (\ref{anhpp}), (\ref{ionpp}), (\ref{dexpp}) and
  (\ref{expp}), as a function of the principal quantum number $n$.
} 
\end{figure}

To get an insight into the magnitude of these interaction probabilities
we show in~\fref{dip} the average values of the annihilation,
ionization, excitation and de-excitation probabilities per unit length.
The average is taken over the even $z$-parity states (i.e., $l-m$
even) for fixed $n$.  The atoms are created in even $z$-parity states
($l=m=0$) and the transitions to odd $z$-parity ones are strongly
suppressed. The figure shows the probabilities using the 
coherent (interaction with the atom as a whole) contribution of the
\textit{Born2} set of cross sections.  This cross section set will be
described in~\sref{crsecc}.  Any other choice among the cross
section sets described in~\sref{crsecc} would lead to very similar results. 
The averages are defined as
\begin{eqnarray}
  \label{anhpp}
\overline{p}_{n}^{\mathrm{anh}}
  = \displaystyle\frac{1}{n(n+1)/2}\sum_{lm} p_{nlm}^{\mathrm{anh}}\;, \\
  \label{ionpp}
                                %
\overline{p}_{n}^{\mathrm{br}} 
  = \displaystyle\frac{1}{n(n+1)/2}\sum_{lm} p_{nlm}^{\mathrm{br}}\;, \\
  \label{dexpp}
\overline{p}_{n}^{n^{\prime}<n}
  =
  \displaystyle\frac{1}{n(n+1)/2} \sum_{lm} 
  \sum_{n^{\prime}<n,l^{\prime}m^{\prime}}
    p_{nlm}^{n^{\prime}l^{\prime}m^{\prime}}\;, \\
  \label{expp}
\overline{p}_{n}^{n^{\prime}>n}
  =
  \displaystyle\frac{1}{n(n+1)/2}
  \sum_{lm} \left( p_{nlm}^{\mathrm{em}} - p_{nlm}^{\mathrm{br}} -
        \sum_{n^{\prime}
          \leq n,l^{\prime}m^{\prime}}
        p_{nlm}^{n^{\prime}l^{\prime}m^{\prime}}\right)\;, 
\end{eqnarray}
where $n(n+1)/2$ is the number of even $z$-parity states for a given $n$.

\subsection{Pionium Evolution in the Target}

To 
simulate the evolution of a pionic atom
we use  the following algorithm:
\begin{enumerate}
\item We generate a laboratory momentum $\vec{P}$, an initial set of
  quantum numbers and an initial position $\vec{R}$ for the atom as
  described in subsection~\ref{creation}.
\item We generate a free path according to:
  \begin{equation}
    p(x) \,\rmd x = \frac{1}{\lambda_{nlm}} \rme^{-x/\lambda_{nlm}}\,\rmd x
  \end{equation}
  where $\lambda_{nlm} = (1/\lambda_{nlm}^{\mathrm{em}} + 1/\lambda_{nlm}^{\mathrm{anh}})^{-1}$
  is the mean free path before either an electromagnetic interaction or the
  annihilation takes place.
\item We displace the atom by the distance $x$:
  \begin{equation}
    \vec{R}^{\prime} = x \frac{\vec{P}}{P} + \vec{R}.
  \end{equation}
\item We determine whether the atom has been annihilated, excited (or
  de-excited) in a discrete collision, or broken up. The relative
  weights of the respective branches of the evolution 
  are given by the probabilities of
  equations~\eref{pannih},~\eref{pdisc}, and~\eref{pbreak}. 
\item If the atom has been scattered and suffered a discrete
  transition we return to step (ii) using the new quantum numbers
  $n^{\prime}$, $l^{\prime}$ and $m^{\prime}$ and the new position
  $\vec{R}^{\prime}$ as the initial values. 
\end{enumerate}

More details on this model may be found in~\cite{sant}.

\section{Break-up Probability Calculation}
\label{pbrs}

In principle, the break-up probability calculation
of pionium should be straightforward once we have established the Monte
Carlo model. The rest would be a matter of generating an atom sample
and computing how many of them break up in the target. However,
two main difficulties arise when trying to implement the
algorithm. 

The first difficulty is due to the presence of an infinite number of
atomic bound states in the calculations. Clearly, only a finite number
of states can be taken into account in the simulation of the evolution
of pionium.  In our calculations we have imposed a cut on the states
with $n \leq n_{\max}$. This would not pose a serious problem if the
atoms, being created mainly in very low $n$ states, could not get 
highly excited. Unfortunately, excitation to ever higher lying bound
states constitutes a major branch in the evolution of pionium.
As a consequence we cannot directly calculate the break-up
probability as outlined in the previous paragraph.

The other difficulty lies in the fact that for some of the cross 
section sets to be 
studied in~\sref{crsecc}, the break-up cross sections have not
been calculated. In this case it is imperative to find  an indirect
way to compute the break-up probability. 

We have discussed in the previous section that pionium terminates 
its evolution in the target by being either annihilated or
broken up. However, the atom can also leave the target in a bound
state. This would happen if one of the generated free paths in the
Monte Carlo procedure carries it to a position outside the target. The break-up
probability ($P_{\mathrm{br}}$), the annihilation probability 
($P_{\mathrm{anh}}$), and
the probability to leave the target in a discrete state ($P_{\mathrm{dsc}}$)
are related by: 
\begin{equation}
\label{prcomp}
1 = P_{\mathrm{br}} + P_{\mathrm{anh}} + P_{\mathrm{dsc}}\;. 
\end{equation}
This equation  allows us to compute the
break-up probability indirectly. 

\subsection{Computation Difficulties Due to Physical Characteristics
  of the Problem} 
\label{compup}

The probability to generate an atom in a specific shell decreases as
$1/n^3$. This means that the number of atoms created with $n \geq 4$
is very small. If the atoms could not get excited to states with large
$n$, we could safely solve the evolution system by setting $n_{\max} >
4$. However, as we saw in~\fref{dip}, the atoms have a tendency to be
excited, as $n$ increases, rather than being annihilated or ionized.

Hence we expect a significant fraction of atoms excited into 
$n>n_{\max}$ shells even for large values of $n_{\max}$. The
probability of an atom in a $nlm$ state to be excited into a state
beyond the cut, i.e.~with $n>n_{\max}$, is given by 
\begin{equation}
\label{pncont}
p_{nlm}^{n^{\prime}>n_{\max}} =
  p_{nlm}^{\mathrm{em}} - p_{nlm}^{\mathrm{br}} -
        \sum_{n^{\prime}
          \leq n_{\max},\,l^{\prime}m^{\prime}}
        p_{nlm}^{n^{\prime}l^{\prime}m^{\prime}}
\end{equation}
where we have used~\eref{pdisc},~\eref{pbreak} and~\eref{pem}. 
However, once the atom jumps into one of these states we
loose control over it and we have to stop its evolution. 

\begin{figure}
\centerline{\includegraphics[width=10.5cm]{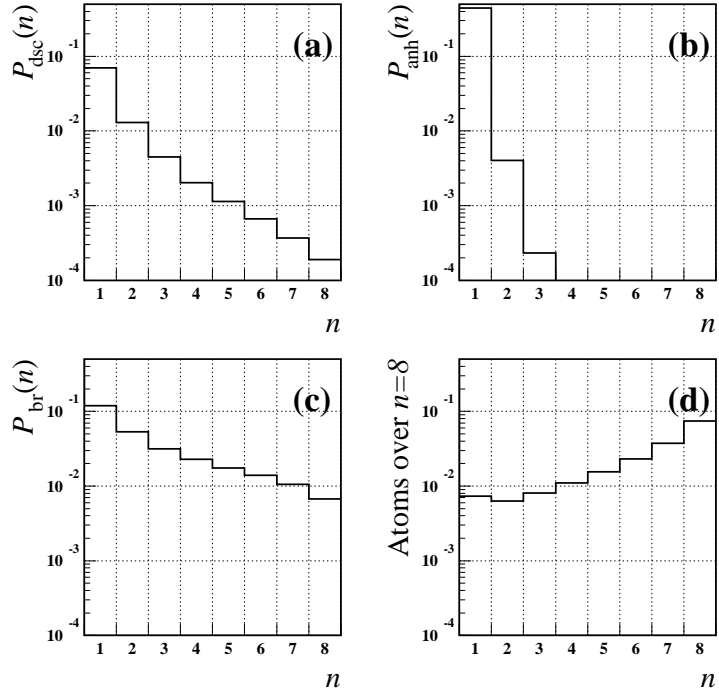}}
\caption{\label{result} Probabilities of finishing the evolution 
  in a discrete state (a), by annihilation (b), or by 
  ionization (c) as a function of the parent state's principal quantum
  number. 
  In (d) we show the probability for an atom in an state $n$  to be
  excited into a non-controlled state with $n>8$. The results are for
  pionium in a 
  $95\;\mu$m~Ni target and the lifetime is assumed to be 
  $3 \cdot 10^{-15}$~s.}
\end{figure}

To analyze the change of the Monte Carlo results with $n_{\max}$ we
have modeled the evolution of a sample of atoms by changing $n_{\max}$
from 7 to 9. We observed three main effects:
\begin{itemize}
\item The fraction of annihilated atoms 
($P_{\mathrm{anh}}(n\leq n_{\max})$) does
  not change significantly. 
\item The portion of atoms leaving the target in discrete states 
  ($P_{\mathrm{dsc}}(n\leq n_{\max})$) changes only slightly.
\item The fraction of dissociated atoms ($P_{\mathrm{br}}(n\leq n_{\max})$)
  changes significantly. 
\end{itemize}
This effect can be understood by checking the dependence of the
annihilation, the discrete and the break-up probabilities on $n$, the
principal quantum number of the state from which the atom was annihilated, 
broken-up, or in which it left the target. In~\fref{result} we show
the result of the Monte-Carlo 
simulation with $n_{\max} =8$ for a sample of one million atoms
using the \textit{Born2} cross section set that also includes cross
sections for the ionization (refer to~\sref{crsecc}).  For the
annihilated atoms we can see that $P_{\mathrm{anh}}(n)$ is negligible for
values of $n \gtrsim 4$. The $P_{\mathrm{dsc}}(n)$ dependence also shows a
fast, but less drastic, decrease with $n$. Only the solution for the
states with $n = n_{\max}-1$ or $n=n_{\max}-2$ is unstable under 
variation of $n_{\max}$. For $n_{\max}=8$ this is a small contribution to the
total $P_{\mathrm{dsc}}$ value. Finally, $P_{\mathrm{br}}(n)$ decreases 
very slowly with
$n$, showing that there is a significant fraction of atoms broken up from
states with $n>n_{\max}$. The probability of an atom to be excited into 
such a state with $n>n_{\max}$ is also shown as a function of the principal
quantum number of the last state before the excitation. 
Obviously, this effect is non-negligible. 

For the cross section sets without break-up cross sections, we can
calculate directly only the total probability for all electromagnetic
processes and the probabilities for discrete transitions to 
states with $n'<n_{\max}$. In these cases we can  therefore not 
distinguish whether an atom has been broken up or excited into a
state with $n'>n_{\max}$, that is, we can only determine the 
\emph{combination} of probabilities  
\begin{equation}
p_{nlm}^{n^{\prime}>n_{\max}}  + p_{nlm}^{\mathrm{br}} =
  p_{nlm}^{\mathrm{em}} - 
        \sum_{n^{\prime}
          \leq n_{\max},\,l^{\prime}m^{\prime}}
        p_{nlm}^{n^{\prime}l^{\prime}m^{\prime}}.
\end{equation}
This is, of course, equivalent to~\eref{pncont}, but in this case 
$p_{nlm}^{\mathrm{br}}$ is unknown.
Thus, for these cross section sets not even the break-up probability
for low $n$ states could be directly calculated and we are forced to use
the procedure described below.

\subsection{Calculation Procedure}

Based on the fast decrease of $P_{\mathrm{anh}}$ and $P_{\mathrm{dsc}}$ as a
function  of $n$ we can assume that almost every atom excited to a
state $n>n_{\max}$ will be eventually broken up. This will be true
even though the excitation probability per unit length 
of a given bound state is significantly larger than the break-up
probability per unit length. We can explain it as follows. The mean
free path of the excited atoms strongly decreases with increasing $n$.
For $n\sim 8$ the mean free path is 
$\lesssim 0.1\;\mu$m.  
An excited atom will thus interact many times within a very short
distance. In every scattering the atom will have some small
probability to break up, thereby terminating its evolution. 
In summary, 
the most probable evolution of an atom that has been excited to
any state with $n \gtrsim 4$  is a sequence of excitations (and 
less frequent de-excitations) terminated by break-up.

However, while we are neglecting the atomic annihilation from states
with $n > 8$ and thus setting $P_{\mathrm{anh}} = P_{\mathrm{anh}}(n \leq 8)$,
 we can
estimate $P_{\mathrm{dsc}}(n > 8)$  by means of a fit to the 
$P_{\mathrm{dsc}}(n)$
histogram as recommended in~\cite{afan}
\begin{equation}
\label{fit}
P_{\mathrm{dsc}}(n > 8) = \frac{a}{n^3} + \frac{b}{n^5}.
\end{equation}

Hence, taking into account~\eref{prcomp} we obtain
\begin{equation}
P_{\mathrm{br}} = 1 - P_{\mathrm{dsc}} - P_{\mathrm{anh}},
\end{equation}
where $P_{\mathrm{dsc}}$ consists of two parts,
\begin{equation}
P_{\mathrm{dsc}} = P_{\mathrm{dsc}}(n \leq 8) + P_{\mathrm{dsc}}(n > 8),
\end{equation}
of which $P_{\mathrm{dsc}}(n \leq 8)$ is computed directly and 
$P_{\mathrm{dsc}}(n > 8)$  is calculated from~\eref{fit}.  
In this manner we can 
calculate the break-up probability even without ionization
cross sections as input. 

\begin{figure}
\begin{center}
\includegraphics[width=7.5cm]{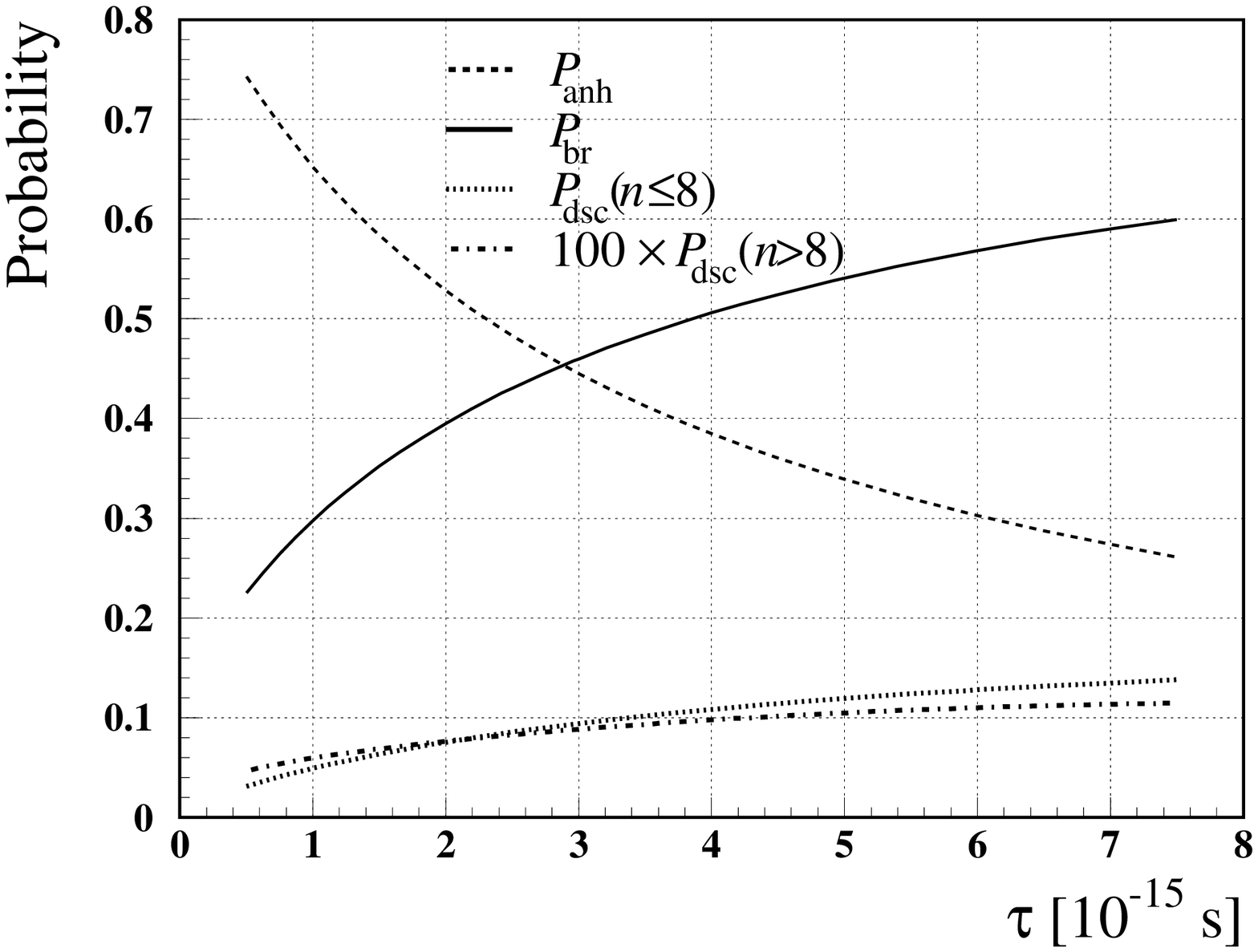}
\includegraphics[width=7.5cm]{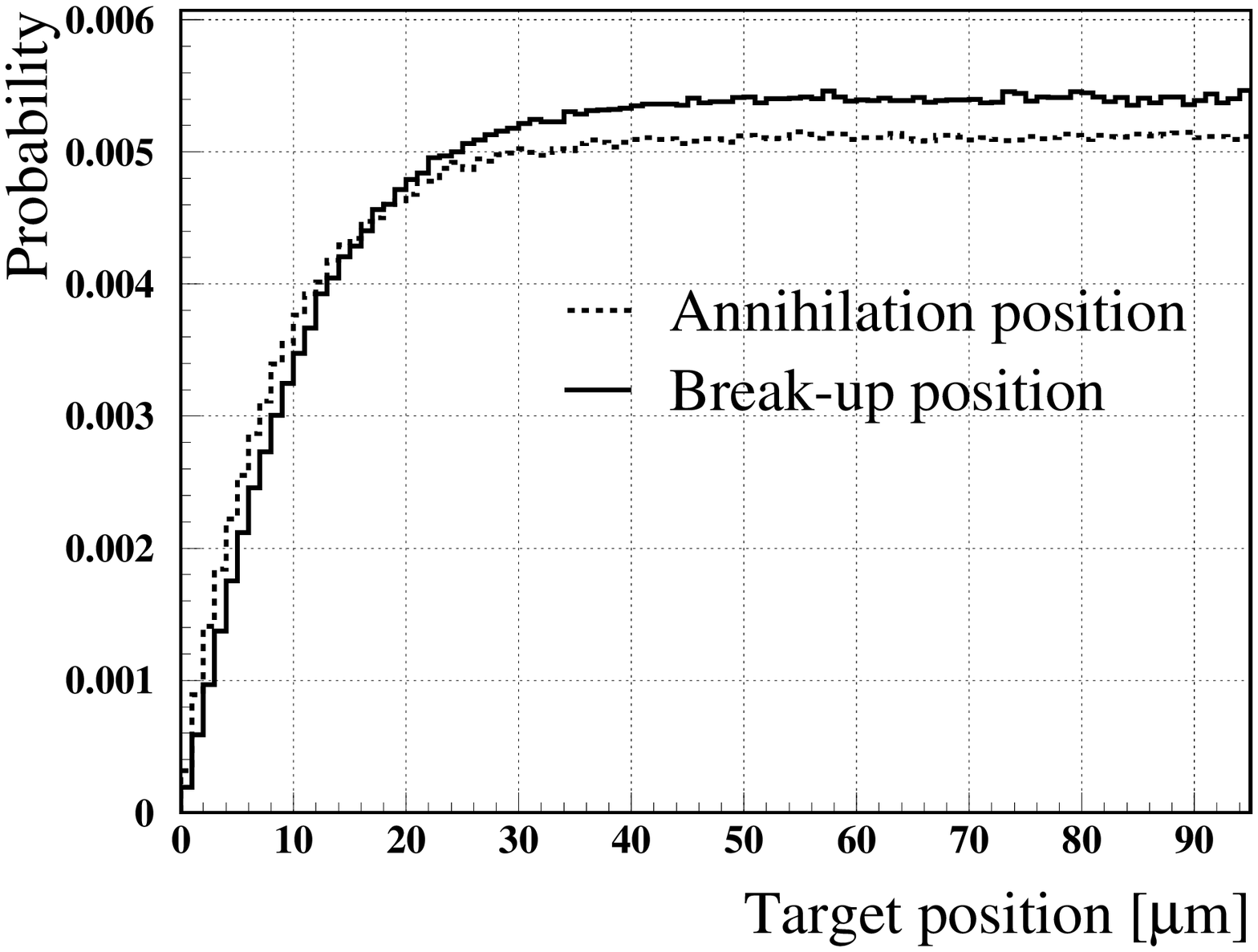} \\
\includegraphics[width=7.5cm]{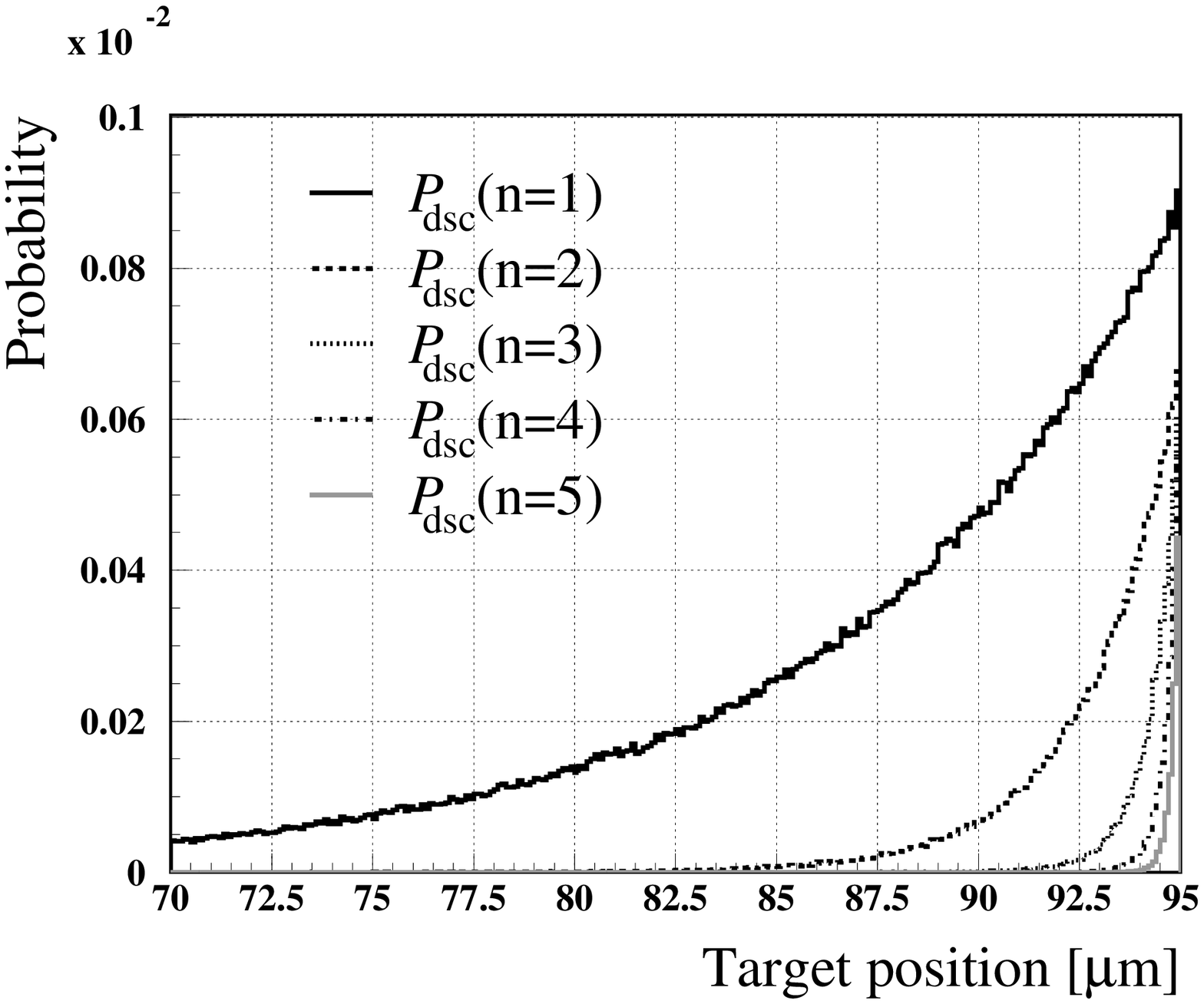}
\end{center}
\caption{\label{probs} 
Top left: The break-up, annihilation, and discrete 
probabilities as a function of lifetime. 
Top right: Break-up and annihilation position distributions.
Bottom: Creation position of those atoms that  leave the target in a
bound state (and contribute to  $P_{\mathrm{dsc}}$). Note that as $n$
increases, only the atoms very near the target end can escape from it. 
All three plots refer to a $95\;\mu$m~Ni target. In the
last two, the lifetime is assumed to be $3\cdot 10^{-15}$\;s. }
\end{figure}

In~\tref{pbbt} and in~\fref{probs} (top left) we show a few
 results for the probability for different lifetime values in 
a $95\;\mu$m~Ni target. The target choice coincides with that of the DIRAC
experiment. We observe that the result of $P_{\mathrm{dsc}}(n>8)$ adds only
a small correction. In~\fref{probs} we also show the ionization and
annihilation distributions as a function of the target position, and
finally the creation position for those atoms that managed to emerge from 
the target in a bound state. As emphasized in subsection~\ref{compup},
with increasing $n$ only the atoms very near the target end will be
able to leave the target in a discrete state. 

\begin{table}
\begin{center}
\caption{\label{pbbt} Results for the different probabilities defined
  in~\eref{prcomp}, as calculated with the \textit{Born2} cross section set
  for a sample of ten million pionic atoms in a $95\;\mu$m thick
  Nickel target.} 
\begin{tabular}{ccccc}
\hline
$\tau [10^{-15}\;\textrm{s}]$ 
& $P_{\mathrm{br}}$ 
& $P_{\mathrm{anh}}$ 
& $P_{\mathrm{dsc}}(n\leq8)$ 
& $P_{\mathrm{dsc}}(n>8)$ \\
\hline
1 & 0.2976 & 0.6527 & 0.0491 & 0.0006 \\
2 & 0.3951 & 0.5287 & 0.0754 & 0.0008 \\
3 & 0.4599 & 0.4451 & 0.0941 & 0.0009 \\
4 & 0.5062 & 0.3848 & 0.1080 & 0.0010 \\
5 & 0.5408 & 0.3392 & 0.1190 & 0.0010 \\
6 & 0.5681 & 0.3029 & 0.1279 & 0.0011 \\
7 & 0.5901 & 0.2740 & 0.1348 & 0.0011 \\
\hline
\end{tabular}
\end{center}
\end{table}

\section{Cross Sections Sets}
\label{crsecc}

In our calculations of the break-up probability we employed three
different sets of cross sections. The first two have been calculated
in the framework of the Born approximation.  We assign the labels 
\textit{Born1} to the calculations made in reference~\cite{afan} and 
\textit{Born2} to those of~\cite{heim}. The two sets differ in four main
points:  
\begin{itemize}
\item The \textit{Born1} set neglects the contribution of incoherent
  scattering (collisions leading to an excitation of
  the target atom), thus considering the coherent interaction only
  (collisions with the target atom as a whole), i.e.~the leading
  term. By contrast the \textit{Born2} set accounts for target 
  excitations.
\item The \textit{Born1} set uses Moli\`ere's  parameterization 
  \cite{moliere} for the
  Thomas-Fermi equation solution as the target atom form factor of the
  pure electric interaction, whereas the \textit{Born2} set takes 
  electron orbitals determined numerically within the Hartree-Fock 
  framework 
  for the same purpose. The Thomas-Fermi-Moli\`ere parameterization of
  the atomic form factor is accurate for low momentum exchange, but
  gives a  small excess for harder scattering. 
\item The \textit{Born1} set considers the sudden approximation (no 
  recoil energy for the target and the pionic atom) and neglects the
  energy difference between the initial and the final state, while
  the \textit{Born2} set accounts for these two effects.
\item Finally, \textit{Born2} set also considers the effect of magnetic
  and relativistic terms. 
\end{itemize}

\begin{figure}
\begin{center}
\includegraphics[width=5cm]{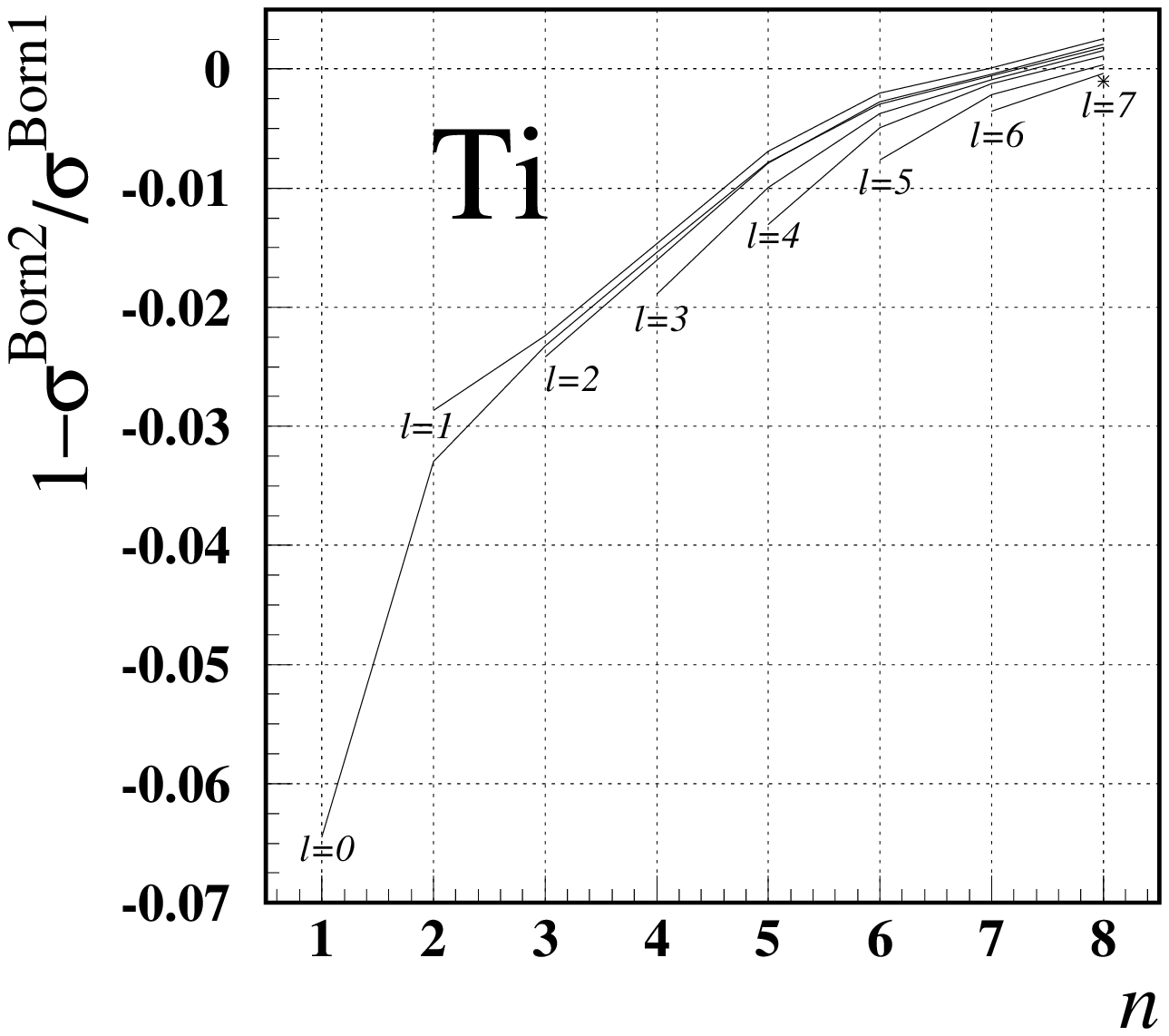}
\includegraphics[width=5cm]{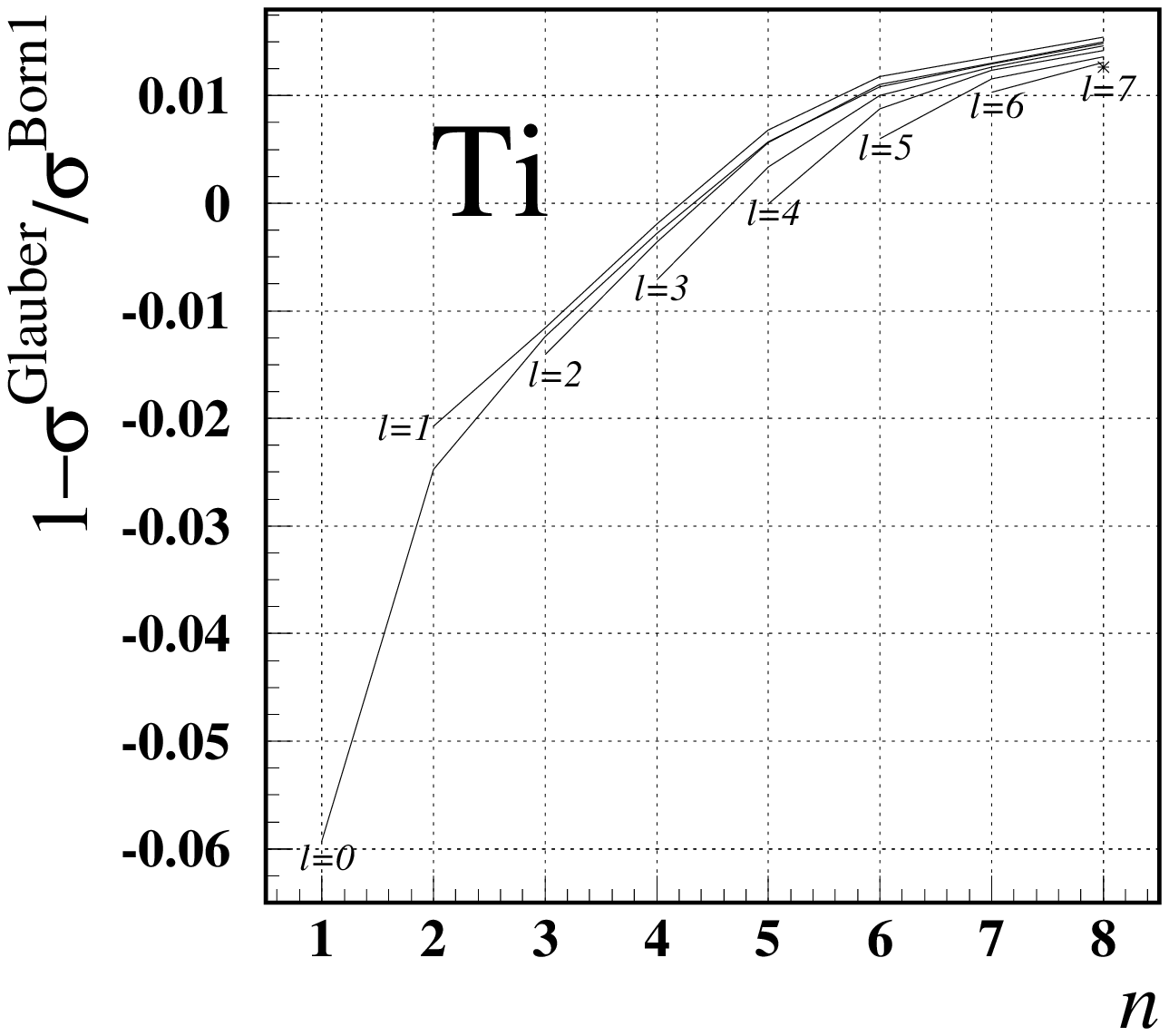}
\includegraphics[width=5cm]{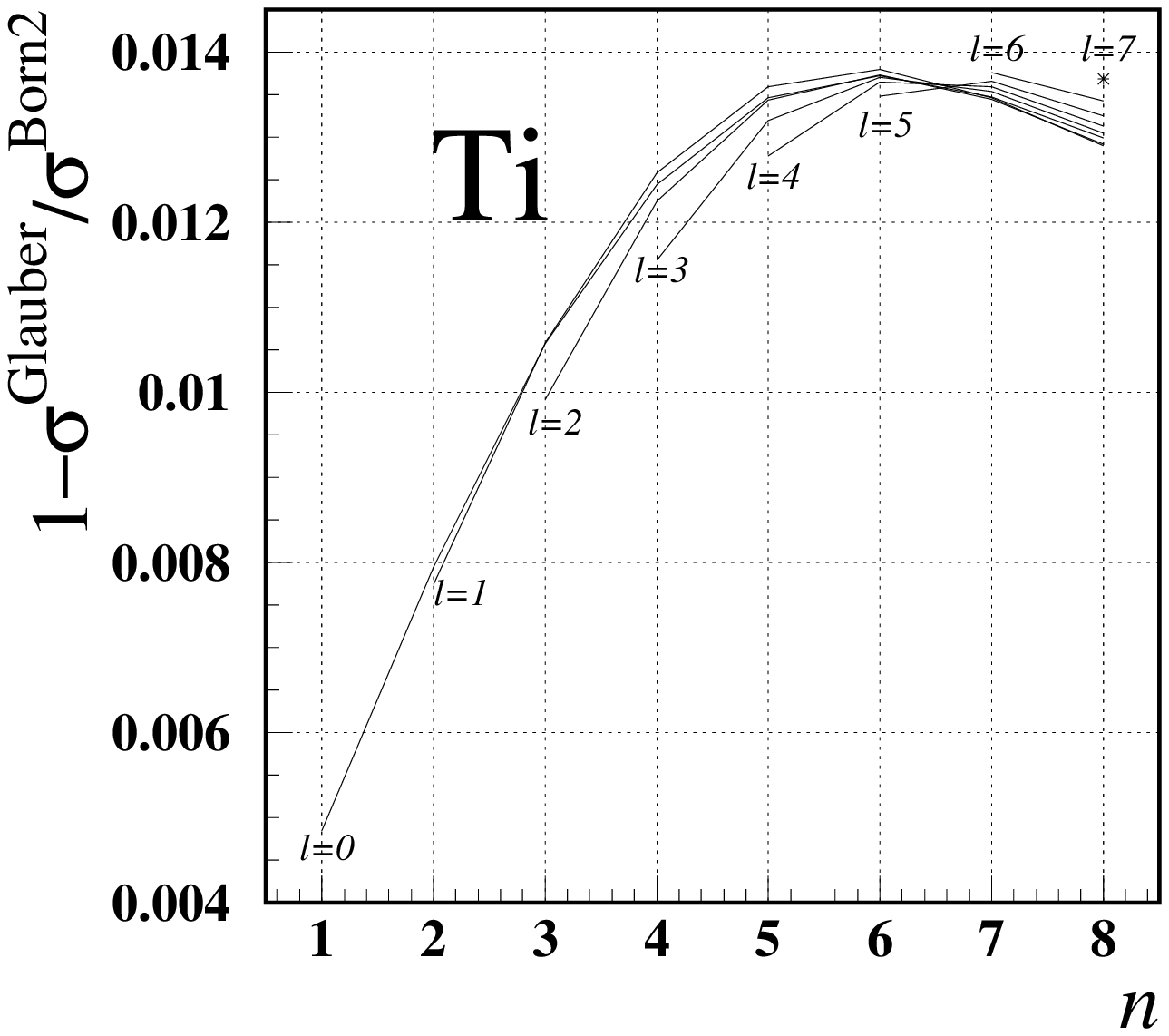}
\includegraphics[width=5cm]{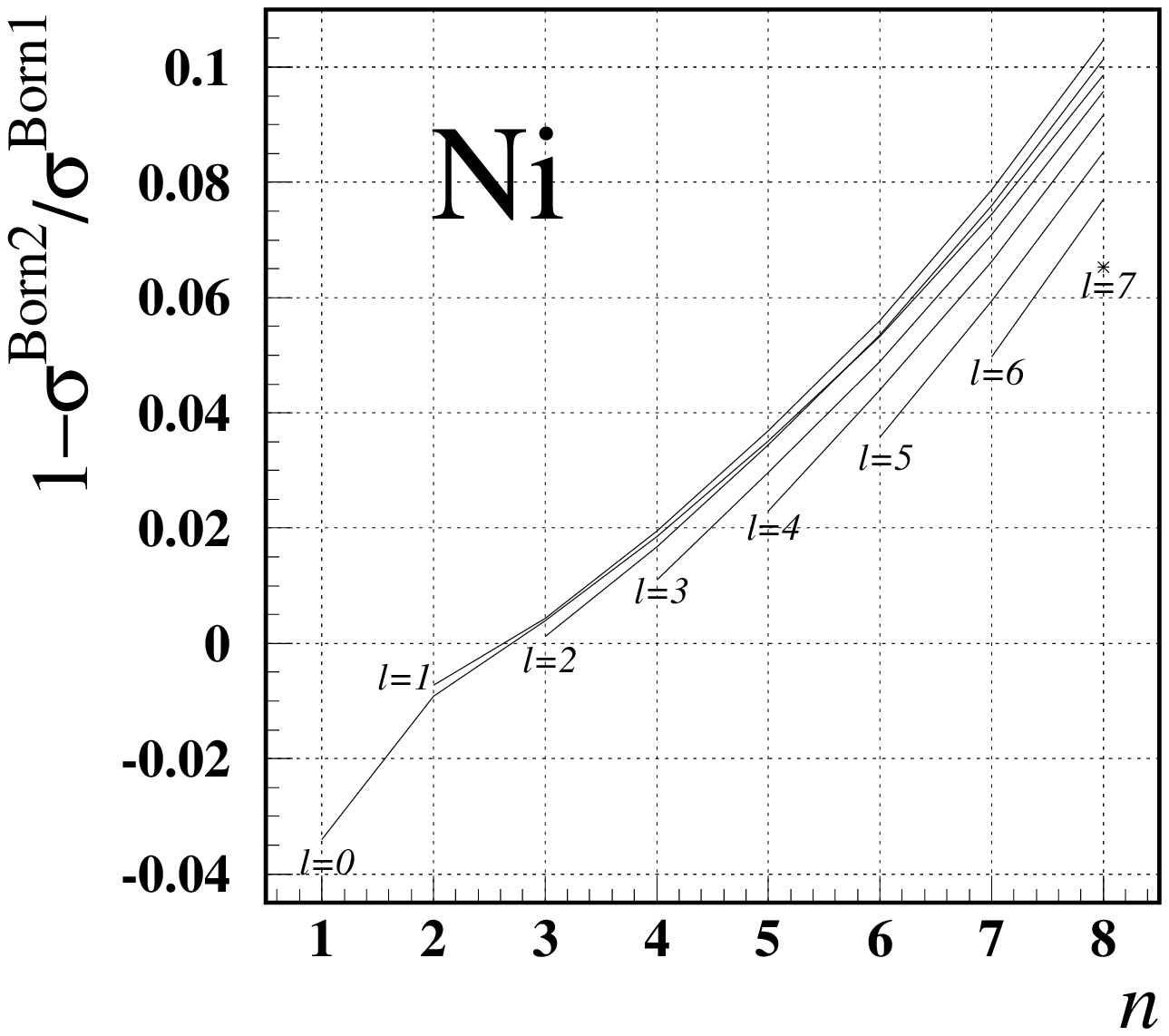}   
\includegraphics[width=5cm]{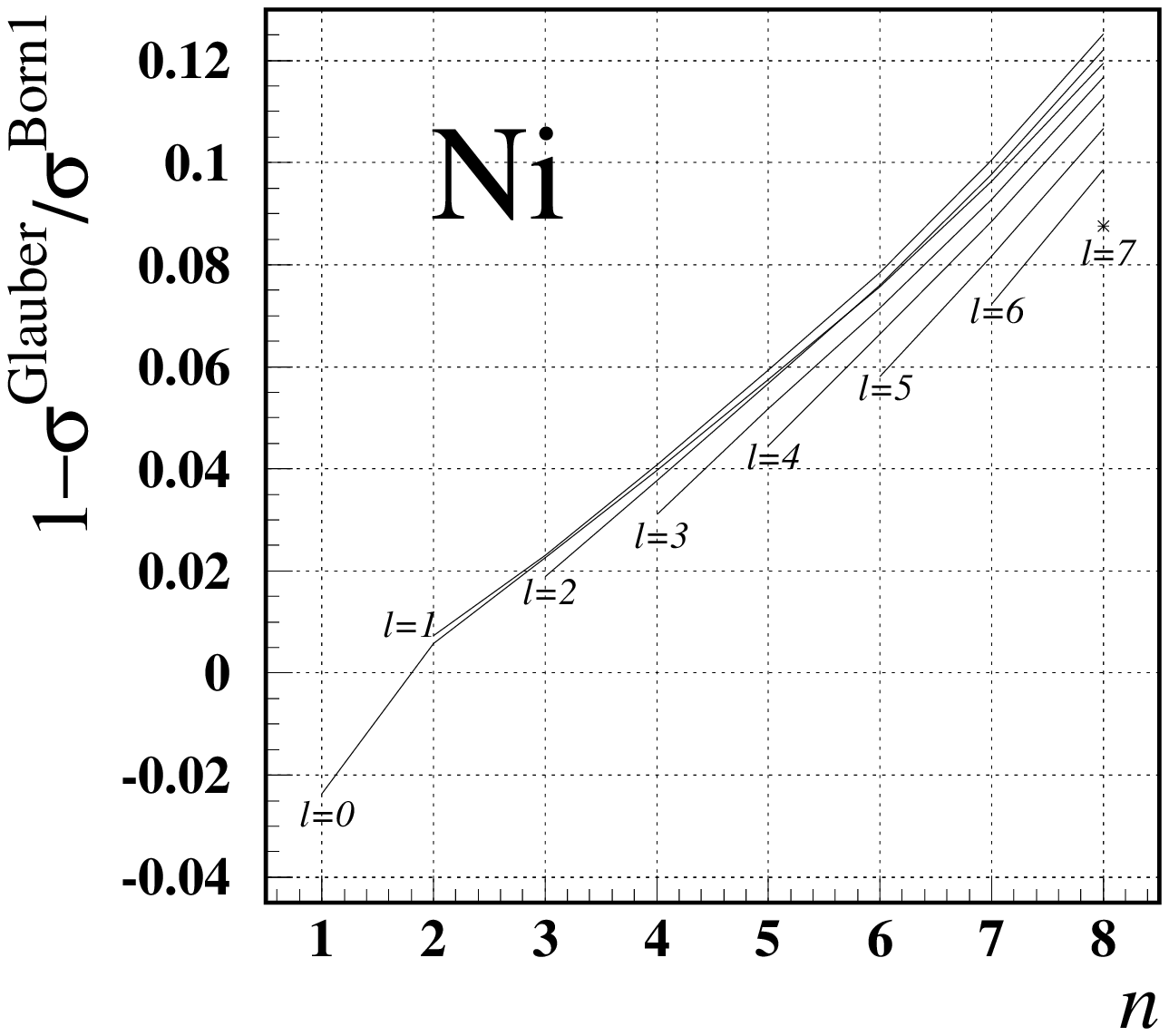}   
\includegraphics[width=5cm]{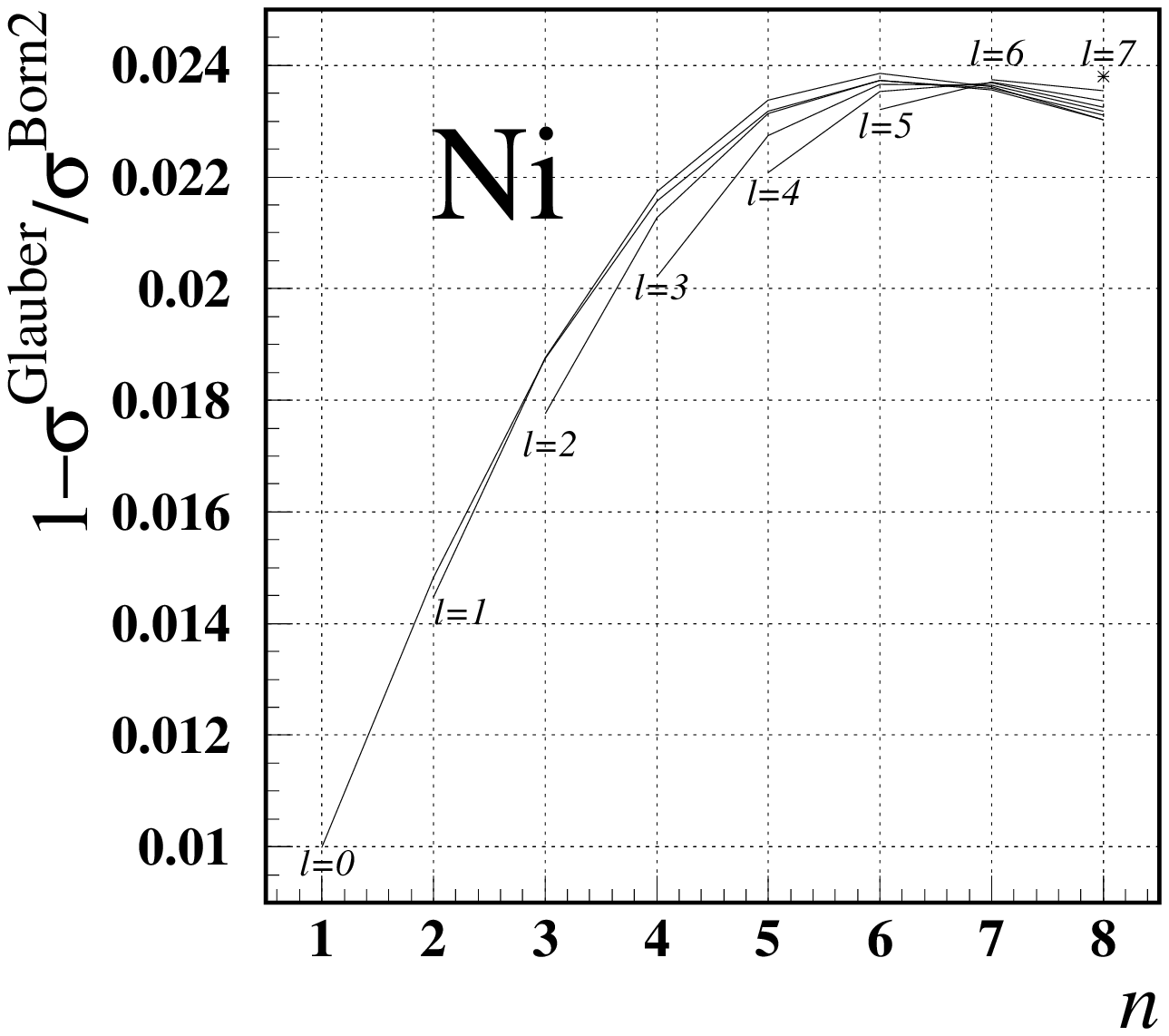}  
\includegraphics[width=5cm]{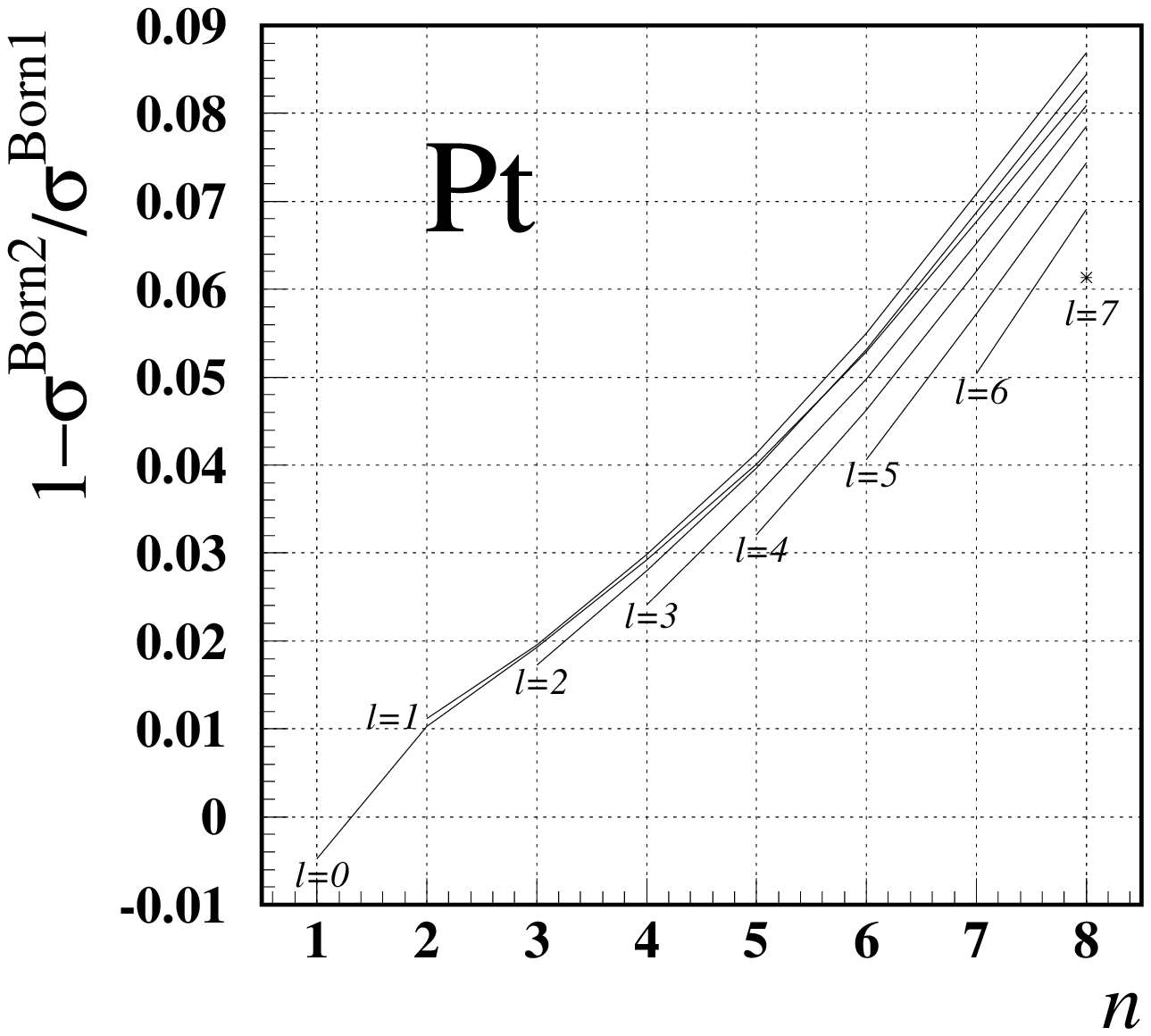} 
\includegraphics[width=5cm]{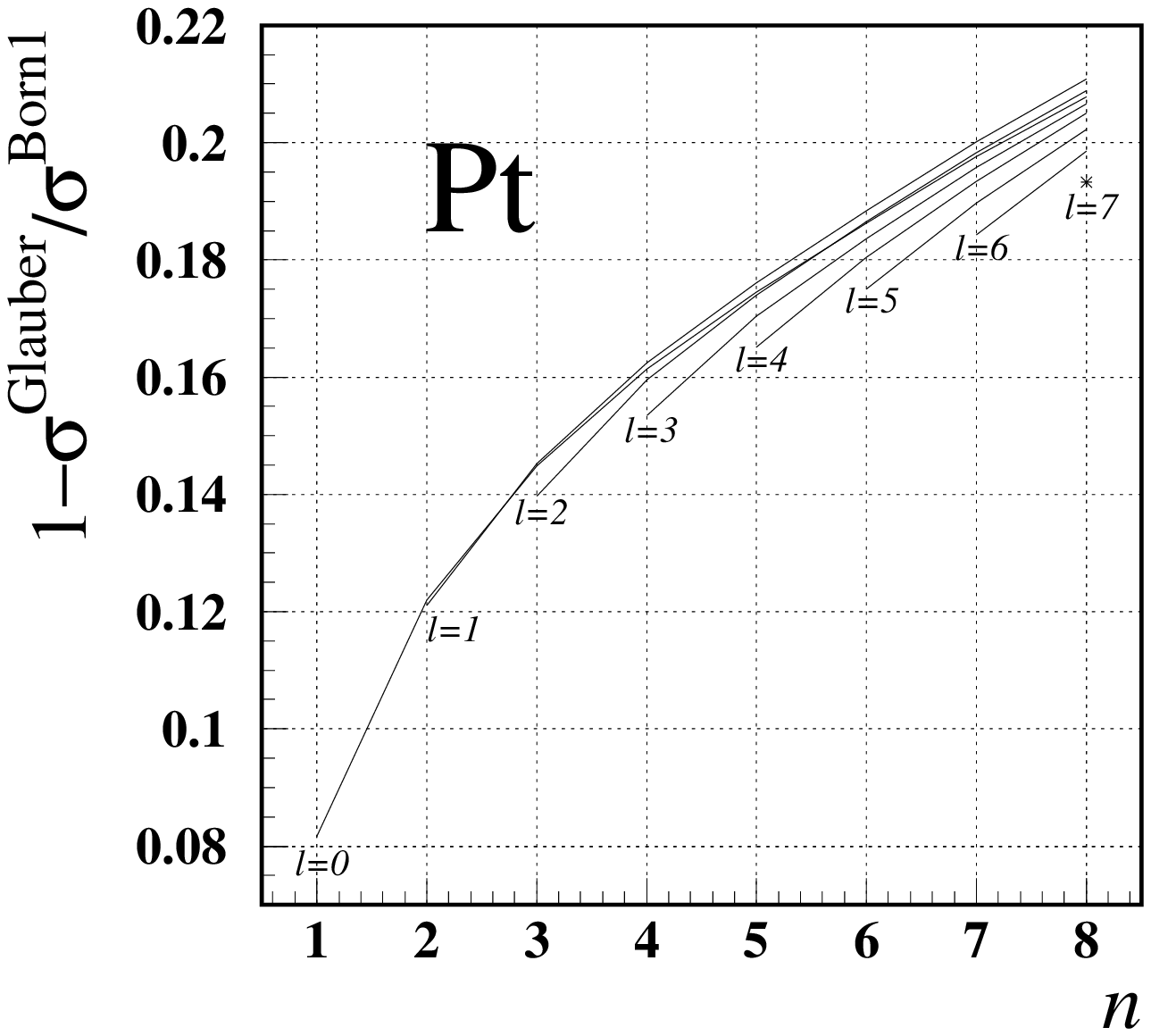} 
\includegraphics[width=5cm]{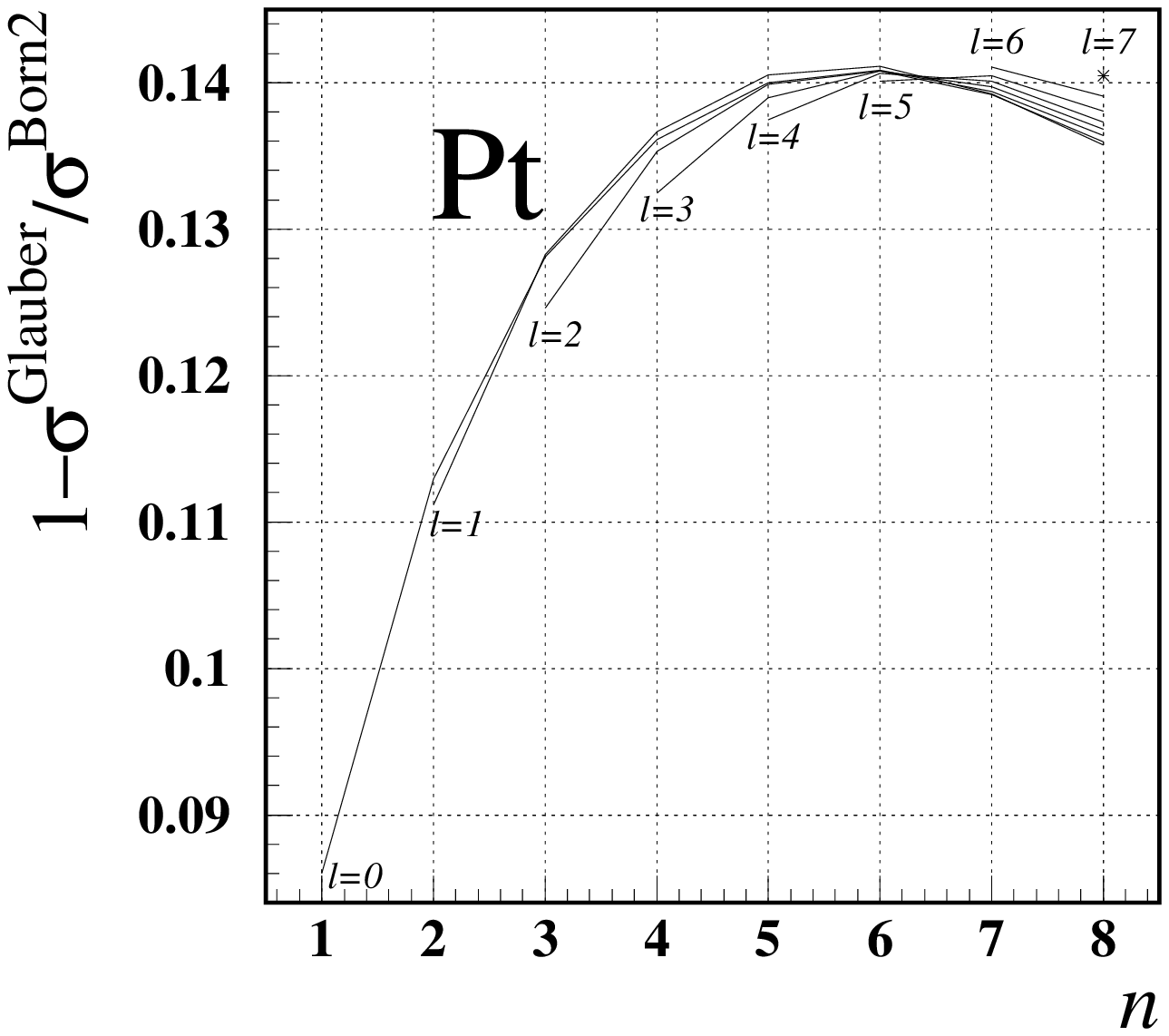}
\end{center}
\caption{\label{rate} 
  In the left column we compare the \textit{Born1} and \textit{Born2}
  cross section sets. The middle column shows a comparison of 
  the \textit{Born1} and
  \textit{Glauber} sets and finally on the right we compare the \textit{Born2}
  and \textit{Glauber} sets. The comparison is made for Titanium
  ($Z=22$), Nickel ($Z=28$) and Platinum ($Z=78$). The plots refer to
  total electromagnetic cross sections averaged over $m$ for even 
  $z$-parity states.
} 
\end{figure}

In principle it has been concluded~\cite{heim} that accounting for 
second order effects like the magnetic terms of the Hamiltonian, the
recoil energy of the atoms, or the relativistic terms generally leads to an
overall decrease of the sudden approximation pure electrostatic
coherent cross section value due to destructive interference with the
leading orders. Moreover,  employing atomic orbitals obtained in
the Hartree-Fock approach for the 
form factor used to compute these cross sections leads to lower values
than those of the \textit{Born1} set since the Moli\`ere parameterization
of the solution to the Thomas-Fermi equation is excessive for mean and
large values of the photon  momentum transfer.  This last issue leads
to discrepancies that increase for large $n$ states and decrease for
large $Z$ target atoms. The difference appears to be balanced 
by neglecting the
incoherent contribution to the cross section in the \textit{Born1}. This
results in a systematically smaller ground state cross section of 
\textit{Born1} set whereas for larger $n$ \textit{Born1} cross sections are
larger (up to $\sim 10\%$ discrepancy) or compatible with \textit{Born2}
results. We can observe the comparison for three target materials
in~\fref{rate} and we shall analyze the disagreement in the
break-up probability resulting from this effect. 

Finally we have also used a set of cross sections where the Glauber
formalism has been applied to calculate the coherent contribution to
the cross section value. The details are given in~\cite{schu}. This
calculation technique accounts for multi-photon exchange in the
pionium--target atom collision. Contrary to what one would expect the
consideration of more than one photon being exchanged diminishes the values
of the cross sections due to a destructive interference 
of the $n$-photon exchange contributions (this happened also when 
accounting for magnetic terms in the \textit{Born2} set). The leading
order of the \textit{Glauber} result matches the sudden approximation of
the Born cross sections (since both neglect the difference between the
initial and the final state energies). However, this cross section set
uses a parameterization for the target atom form factors similar to the
ones used in 
\textit{Born2}. This explains the disagreement with respect to the
\textit{Born1} and the agreement with \textit{Born2} set for low $Z$ targets,
as can be seen in~\fref{rate}. The corrections due to multiphoton
exchange are important for large $Z$ targets~\cite{schu} and this
explains the large discrepancies obtained for Platinum. 

\section{Results and Conclusions}
\label{results}
After the discussions of the previous sections, we finally present the
results 
of the break-up probability calculation. The DIRAC experiment has the
possibility of choosing between several targets. The design of these
different targets has been made to achieve the  maximum break-up
probability resolution in different lifetime ranges. Large $Z$ targets
with larger interaction cross sections are better suited for small lifetime
values whereas lower $Z$ materials are more sensitive to larger
lifetime values. Three of these targets are the Pt $28\;\mu$m
target, suitable for lifetime ranges $\tau < 1 \cdot 10^{-15}\;$s,  the
Ni $95\;\mu$m target for  
$\tau \sim 3 \cdot 10^{-15}\;$s
and the Ti $251\;\mu$m target for $\tau \sim 4 \cdot 10^{-15}\;$s.
The target thickness was chosen so as 
to have the same radiation length and hence equivalent
multiple scattering effects for all three targets. 
The Nickel target constitutes DIRAC's main target with which 90\% of the
  data have been collected, as it is optimal for the theoretically
  predicted lifetime value.

\begin{figure}
\begin{center}
\includegraphics[width=8cm]{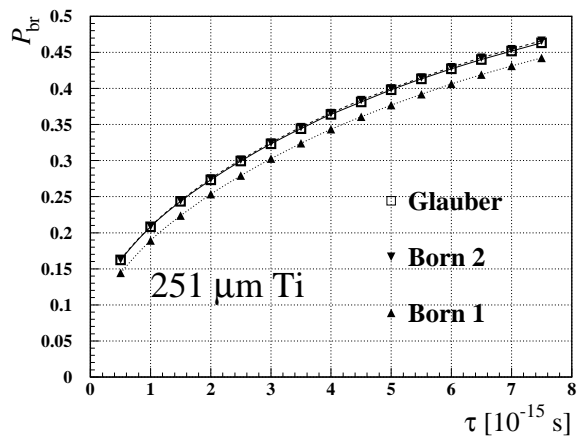}
\includegraphics[width=8cm]{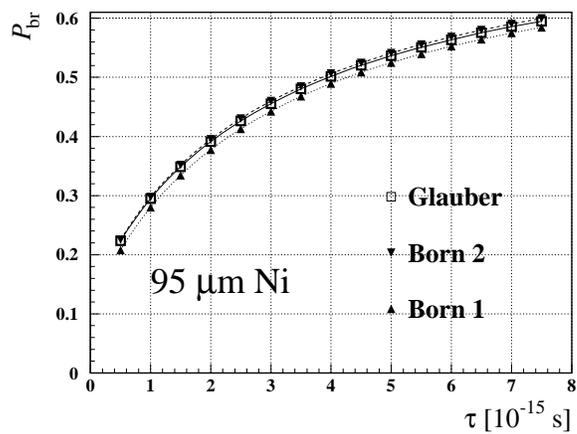}
\includegraphics[width=8cm]{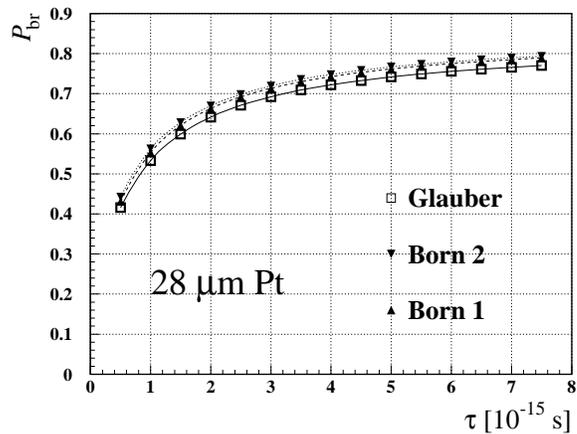}
\end{center}
\caption{\label{pbr} 
  The break-up probability results for the three cross section
  sets and the three target materials.
}
\end{figure}

In~\fref{pbr} the break-up probability curves are shown for these
three targets. The calculation has been carried out for samples of ten
million events, with a statistical error less than $0.08\%$.

One can clearly see that for the Ti and Ni targets the \textit{Glauber} and
\textit{Born2} sets lead to similar results whereas the \textit{Born1} set
shows a $\sim 8\%$ disagreement. For the large $Z$ target (Pt) both
\textit{Born1} and \textit{Born2} are biased toward large values. In this
case, the multi-photon contributions to the total cross sections 
are not negligible. In any case the discrepancies between the break-up
probability results are at the level of the discrepancies between
ground state cross sections and much smaller than the differences
between cross sections sets for medium or highly excited states. We
can understand this based on the fact that the probability for the atoms
to leave the target in a
discrete state
other than the ground state and maybe the first excited
shell  $P_{\mathrm{dsc}}(n \gtrsim 2)$ is of the order of, or smaller than,
5\%. Hence, even large uncertainties in this magnitude (up to $10-15\%$)
lead to very small changes in the break-up probability result. Only
discrepancies in the ground state population and maybe the first
excited shell, where most of the atoms are created, would lead to
significant differences between the break-up probability results of
the different sets. 

Graphically we can view the atom as a balloon being inflated in every
collision with the target. The different sets will lead to similar
results of the size increase rate as long as the atom remains in a low 
excited state. However, as the atom grows (inflates) 
it will no longer be able to advance as easily in the target due to 
its large size 
and will finally break-up (explode). Large discrepancies in the
excitation and break-up rate of the excited atom will not be important
given that the mean free paths for the excited states are very small 
compared to the target dimensions. 

In summary, we recall that the high precision measurement
attempted by the DIRAC collaboration also requires an accuracy to better than $1\%$
in our theoretical break-up probability calculations. We note that the
seemingly large
discrepancies among our different cross section sets particularly 
for pionium transitions starting from highly excited states do not lead to
significant differences in the theoretical break-up probabilities. The
discrepancies between break-up probabilities are coming almost entirely from
differences in the cross sections for the lowest lying states, where both 
the atomic structure of the target and the multi-photon transitions need to
be treated as accurately as possible. This challenge, however, has already
been mastered in our previous works~\cite{heim,afan3} where we showed that the 
required 1\% accuracy can be achieved 
with our calculations, albeit only 
with the \textit{Born2} and the \textit{Glauber} sets for low $Z$ and
with the \textit{Glauber} set for large $Z$ targets.
The important conclusion of the present investigation is the finding that
the (infinitely many!) highly excited states of pionium do \emph{not} 
limit the validity of our approach even though we can include explicitly 
only a moderate number of these states in our simulations.

\begin{table}
\begin{center}
\caption{\label{pbrt} 
  Comparison of the break-up probability results for
  the Ti $251\;\mu$m, the Ni $95\;\mu$m and the Pt 
  $28\;\mu$m targets. The lifetime value is assumed to be
  $3\cdot 10^{-15}\;$s in the calculations.
} 
\begin{tabular}{ccccccc}
\hline
Target 
& $P_{\mathrm{br}}^{\mathit{Born1}}$ 
& $P_{\mathrm{br}}^{\mathit{Born2}}$ 
& $P_{\mathrm{br}}^{\mathit{Glauber}}$ 
& $1-\displaystyle
  \frac{P_{\mathrm{br}}^{\mathit{Born2}}}{P_{\mathrm{br}}^{\mathit{Born1}}}$ 
& $1-\displaystyle
  \frac{P_{\mathrm{br}}^{\mathit{Glauber}}}{P_{\mathrm{br}}^{\mathit{Born1}}}$ 
& $1-\displaystyle
  \frac{P_{\mathrm{br}}^{\mathit{Glauber}}}{P_{\mathrm{br}}^{\mathit{Born2}}}$ 
\\ \hline
Ti & 0.3026 & 0.3249 & 0.3232  & $-7.4\%$ & $-6.8\%$ &  $0.5\%$ \\
Ni & 0.4425 & 0.4599 & 0.4555  & $-3.9\%$ & $-2.9\%$ &  $1.0\%$ \\
Pt & 0.7137 & 0.7196 & 0.6924  & $-0.8\%$ &  $3.0\%$ &  $3.8\%$ \\
\hline
\end{tabular}
\end{center}
\end{table}

\ack

We would like to thank B. Adeva, L.L. Nemenov, L. Tauscher and D. Trautmann
for their support. We are also indebted to M. Pl\'{o}, J.J. Saborido and
A.V. Tarasov for their invaluable help.

\end{document}